\documentclass[fleqn,twoside]{article}
\usepackage{espcrc2}

\usepackage{graphicx,psfrag}

\usepackage[figuresright]{rotating}

\title{\vskip -3.5truecm
       {\normalsize Yamada Conference LVI, the Fourth International
        Symposium on Crystalline Organic Metals,\\
        Superconductors and Ferromagnets\\
        ISCOM 2001 - Abstract Number S67 }
       \vskip 2truecm
       Thermal Conductivity in Vortex State of Nodal Superconductors}
\author{H. Won\address{Department of Physics, Hallym University,
Chunchon, 200-702, South Korea }
                \thanks{HW acknowledges the support by the  Hallum
Academy of Science,Hallym University.
                We thank Y. Matsuda and K. Izawa for
keeping us informed about ongoing experiments on
$\kappa$-(ET)$_2$Cu(NCS)$_2$ and useful discussions. Also
discussions with M. A. Tanatar and Y. Maeno on the c-axis thermal
conductivity  in Sr2RuO4 which initiate our study of the phononic
thermal conductivity are gratefully acknowledged.
We thank also M. A. Tanatar for providing us the digitalized
version of the experimental data from Ref.\cite{Suzuki}, which we used in Fig. 2 and Fig.3.},and
                K. Maki\address{Department of Physics and Astronomy,
        University of Southern California,
        Los Angeles CA 90089-0484, USA}
        \address{Max-Planck Institut f\"{u}r Physik Komplexer Systeme,
        N\"othnitzer Str.38,
        D-01187, Dresden, Germany }}

\begin{document}

\begin{abstract}
How to determine the symmetry  of the superconducting order
parameter  is one of the important issues in  novel
superconductors, which include charge  conjugated organic
superconductors. We have proposed   that the   angular dependence
of the  thermal conductivity in a planar magnetic field provides a
new window to look at the symmetry of the order parameter. After a
brief summary  of the  quasiclassical approach we describe  how
the symmetry of the  superconducting order parameter in
Sr$_2$RuO$_4$, CeCoIn$_5$  and $\kappa$-(ET)$_2$Cu(NCS)$_2$ is
determined. Also in some of experiments the phononic thermal
conductivity plays the crucial role.

\vspace{0.5cm} {\bf Keywords:} Unconventional superconductors,
Thermal conductivity, Vortex state
 \vspace{1pc}
\end{abstract}

\maketitle

\section{Introduction}
Since the   discovery of  the  first organic   superconductor in
(TMTSF)$_2$PF$_6$ (or Bechgaard salts) \cite{Jerome}, the symmetry
in superconducting order  parameter has been one of the important
issues \cite{Takigawa,Hasegawa}. Indeed the symmetry of the
superconducting order parameter becomes one of the central issues
after the establishment of $d$-wave symmetry in both hole and
electron doped high  T$_c$  superconductors
\cite{Harlingen,Tsuei,Maki,Maki_Physica}. Most likely the
superconductivity in (TMTSF)$_2$X  with X = ClO$_4$, PF$_6$, etc.
is of $p$-wave \cite{Maki_Metals}.  In particular, a flat Knight
shift seen in a recent NMR experiment \cite{Lee} is consistent
with this picture. The remaining question  is  whether the p-wave
superconductor belongs   to 1D representation \cite{Hasegawa} or
2D representation \cite{Maki_Metals}. After   a long controversy
\cite{Lang}, $d$-wave superconductivity is emerging in
$\kappa$-(ET)$_2$ salts
\cite{Mayaffre,Nakagawa,Carrington,Pinteric,Arai}. Here now the
question is whether the symmetry  is of d$_{xy}$-wave as suggested
by the theoretical works \cite{Visentini,Schmalian,Louati} or
d$_{x^2-y^2}$-wave as the recent STM study suggests \cite{Arai}.
We shall give a somewhat surprising answer on this based on the
recent angular dependent thermal conductivity data of
$\kappa$-(ET)$_2$Cu(NCS)$_2$ by Izawa et al \cite{Izawa}.
        The quasiparticle spectrum of  all these new superconductors
        is well described
by the BCS theory for nodal (or unconventional) superconductors
\cite{Maki,Maki_Physica,Won}. In particular, there are nodal
excitations (i.e. the quasiparticles which inhabit in the vicinity
of the nodal lines) which persist  to low temperatures (i.e.T$<<
\Delta$ where $\Delta$ is  the superconducting order parameter).
In the vortex state the quasiparticle spectrum  is modified  due
to the supercurrent circling around individual vortices. In order
to describe the quasiparticle spectrum in the vortex state,
Volovik \cite{Volovik} has introduced a very simple method to
evaluate the effect of the supercurrent within the quasiclassical
approximation. In particular,   he has  shown that   the specific
heat in   the vortex  state  in nodal superconductors is
proportional to $\sqrt{H}$   for $H/H_{c2}<<1$ where H  is the
magnetic field strength. This $\sqrt{H}$ dependence has been seen
in YBCO \cite{Moler,Moler_PRB,Revas},  LSCO \cite{Chen},
$\kappa$-(ET)$_2$ salts \cite{Nakagawa} and Sr$_2$RuO$_4$
\cite{Nishizaki,Won_Euro}. We shall show later that in the
presence of impurity the above dependence may change as
\cite{Won_cond}
\begin{eqnarray}
\lefteqn{\frac{C_s}{\gamma_NT}  = \frac{N(0)}{N_0} \{1 + \frac{\Delta}{\pi\Gamma} <x^2>\}
\quad\quad\quad\quad\quad }\nonumber\\
\lefteqn{=     \!\frac{N(0)}{N_0} \{1 +
 \frac{3\tilde{v}^2(eH)}{16\pi\Gamma\Delta} [ \ln(\sqrt{
\frac{3}{2(eH)}} \frac{\Delta}{\widetilde{v}} ) - \frac{1}{27} ]
\}}
\end{eqnarray}
for  Sr$_2$RuO$_4$ in H$//$ a-b. Similarly for $d$+$s$-wave in
H$//$ a-b,
\begin{eqnarray}
\lefteqn{\frac{C_s}{\gamma_NT} = \frac{N(0)}{N_0} \{ 1 +
\frac{\widetilde{v}^2(eH)}{8\pi\Gamma\Delta} [ ( 1 -
\frac12r\cos(2\theta) ) \times }\nonumber\\
\lefteqn{\ln(\frac{\Delta}{\widetilde{v}\sqrt{eH}})
 - \frac{1}{16}( 1 - \frac12( 1 - 2r^2 )\cos(4\theta) ) ] \} }
\end{eqnarray}
Here $N(0)$ is the residual density of states in the presence
of the impurity scattering and $N_0$ is the one for the normal
state.
Also $\gamma_N$ is the Sommerfeld coefficient, $\widetilde{v}=\sqrt{v_av_c}$
 and $v_a$ and $v_c$ are the Fermi
velocities in the a-b  plane and parallel to the c-axis, respectively. For
$d+s$-wave we took $\Delta({\bf k}) \propto \cos(2\phi) - r$.
$\theta$ is the angle ${\bf H}$ makes from the $a$ axis.
 Here $x=
{\bf v} \cdot {\bf q} /\Delta$ and ${\bf v} \cdot {\bf q}$ is
        called the Doppler shift,
where $\bf{v}$ is  the Fermi velocity and 2$\bf{q}$ is the  pair
momentum and $< ... >$ means space average over the vortex lattice
and $\bf{v}$ average over the nodal lines. Some of these details
are given in \cite{Won_press}. In the above derivation, we have
assumed that the system is in the clean limit ( $\epsilon$, $T$
$<< \sqrt{\Gamma\Delta}$ ) while Volovik's result applies for the superclean
limit ($\sqrt{\Gamma\Delta}$ $ \ll$  $\epsilon$, $T$). Here $\epsilon =
\frac12\tilde{v}\sqrt{eH}$ is the characteristic magnetic
energy. Also when we put $r$ = 0 in Eq.(2), we obtain the usual
expression for $d$-wave superconductors. The specific heat in the
clean limit in LSCO in ${\bf H} \parallel {c}$ has been reported
recently\cite{chang}.
The  quasiclassical   approximation
is   extended
 to  calculate   the  thermal conductivity in the vortex  state
\cite{Won_cond,Barash,Kbert,Vekhter,Won_CPA}. Indeed we can now
describe the  angular dependent thermal conductivity observed in
single crystals of YBCO \cite{Salamon,Yu,Aubin,Oca} consistently
if we assume that the system is in the superclean limit and T$>>
\epsilon$ \cite{Won_CPA}.    In the following we shall first
review the theory limiting ourselves to the  clean limit
($\epsilon << \sqrt{\Delta\Gamma}$). As to  the result for the superclean limit
readers may consult \cite{Won_Euro,Won_CPA,Won_Kluwer}.
Also we consider  the phononic thermal  conductivity both
         in Sr$_2$RuO$_4$ and  in $\kappa$-(ET)$_2$
salts. The c-axis thermal conductivity in Sr$_2$RuO$_4$
\cite{Tanatar} appears to be  described in terms of the phononic
thermal conductivity \cite{Won_Maki}.        As to
$\kappa$-(ET)$_2$Cu(NCS)$_2$ it appears the phononic thermal
conductivity dominates for T $>$ 0.5K \cite{Izawa,Won_Maki}. For T
$<$ 0.47K there   appears a  clear sign of the electronic
contribution. From the  angle dependence of the thermal
conductivity, we can deduce $d$+$s$-wave ($\Delta({\bf{k}})
\propto \cos(2\phi) -0.067$) for $\kappa$-(ET)$_2$Cu(NCS)$_2$. In
Fig.1 we show $|\Delta({\bf k})|$ for superconductors a)$s$-wave,
(b) $d$-wave as in HTSC, CeCoIn$_5$\cite{Izawa_PRL},
$\kappa$-(ET)$_2$ salts\cite{Izawa}, (c)2D $f$-wave as in
Sr$_2$RuO$_4$\cite{Nishizaki,Won_Euro} and (d)$f$-wave as in UPt$_3$
\cite{lussier,tou,Maki_Euro}.

\begin{figure}

     \includegraphics[width=80mm]{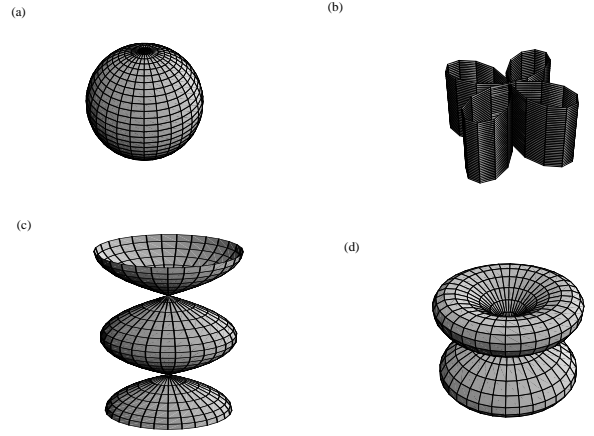}
     \caption{$|\Delta({\bf k})|$'s for $s$-wave and some of unconventional
superconductors are  shown:
(a) $s$-wave superconductor, $\Delta({\bf k})=\Delta$
(b) $d_{x^2-y^2}$-state, $\Delta({\bf k}) \sim \cos(2\phi)$
(c) 2D $f$-wave, $\Delta({\bf k}) \sim\cos(ck_z)e^{\pm \imath\phi}$
(d) $E_{2u}$-state, $\Delta({\bf k}) \sim \cos\theta\sin^2\theta e^{\pm\imath\phi}$ }
\vspace{-0.2cm}
 \end{figure}

\section{Thermal conductivity in the vortex state.}
In the following we limit  ourselves to nodal superconductors;
2D $f$-wave with $\Delta({\bf{k}}) \propto \cos(ck_z)e^{\pm i\phi}$
as in Sr$_2$RuO$_4$ and  $d+s$-wave with $\Delta({\bf{k}}) \propto
\cos(2\phi) - r$ as in $\kappa$-(ET)$_2$Cu(NCS)$_2$. Also we limit
ourselves to the clean limit(i.e. $\sqrt{\Delta\Gamma} \gg \epsilon=
\tilde{v}\sqrt{eH}$, where $\tilde{v}=\sqrt{v_a
v_b}$)\cite{Won_Euro}. Then the thermal conductivity for $T \ll \Delta$ and in a
planar magnetic field is given by
\begin{equation}
\frac{\kappa_{xx}}{\kappa_0 }\!\!=\!\!\!1 +
\frac{\tilde{v}^2(eH)}{4\pi\Gamma\Delta}
\ln({4\sqrt{\frac{2\Delta}{\pi\Gamma}}})
(\ln(\frac{\Delta}{\tilde{v}}\sqrt{\frac{2}{3eH}} ) - \frac{1}{72}
)
\end{equation}
for 2D $f$-wave.
\begin{eqnarray}
\frac{\kappa_{xx}}{\kappa_0 }&=&1 +
\frac{\widetilde{v}^2(eH)}{6\pi\Gamma}
\ln({4\sqrt{\frac{2\Delta}{\pi\Gamma}}})(( 1-\frac12 r
\cos(4\theta)) \times \nonumber \\
& &
\!\!\!\!\!\!\!\!\!\!\!\!\ln{\frac{2\Delta}{\widetilde{v}\sqrt{eH}}}
-\frac1{16}(1-\frac12(1-2r^2)\cos(2\theta)))
\end{eqnarray}
for $d+s$-wave.
\par
Similarly the Hall thermal conductivity is given
by
\begin{equation}
\frac{\kappa_{xy}}{\kappa_0 } =
-\frac{\widetilde{v}^2(eH)}{24\pi\Gamma\Delta}~\sin(2\theta)~
\ln(2\sqrt{\frac{2\Delta}{\pi\Gamma}}~)
\ln(\frac{\Delta}{\widetilde{v}}\sqrt{\frac{2}{3eH}} )
\end{equation}
for 2D $f$-wave, and
\begin{eqnarray}
\frac{\kappa_{xy}}{\kappa_0 }& =&
-\frac{\widetilde{v}^2(eH)}{16\pi\Gamma\Delta} ( 1-r^2
)\sin(2\theta) \ln( 4\sqrt{\frac{2\Delta}{\pi\Gamma}} )\times
\nonumber\\ & &\ln( \frac{\Delta}{\widetilde{v}\sqrt{eH}} )
\end{eqnarray}
for $d+s$-wave.
Here $\kappa_0$ is the thermal conductivity in the absence of the
magnetic field.
\begin{figure}
\vspace{-3cm}
     \includegraphics[width=70mm]{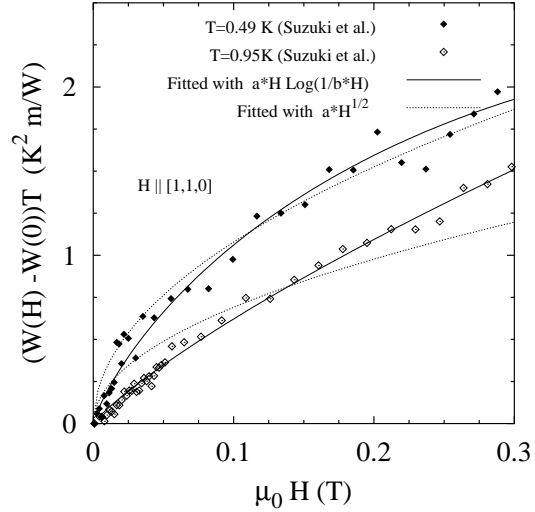}
     \caption{The field dependence of the thermal resistance in the magnetic field
     along [110]direction within the plane Ref.\cite{Suzuki} is compared
     with the fitted
     $a H*log(1/b*H)$ and $\sqrt{H}$.
     }
 \end{figure}

\section{Phononic thermal conductivity}
        So far we have considered only  the electronic thermal conductivity. In general,
the thermal conductivity is written  as
$\kappa=\kappa^{el}+\kappa^g$ , where the second term is due to
phonons. The importance of $ \kappa^g$ in high T$_c$ cuprate
superconductors, YBCO and Bi2212 have  been well-documented
\cite{Chiao}. In low temperatures (for T $<$ 5K ) $\kappa^g \sim
T^3$ in single crystals of YBCO and Bi2212 \cite{Chiao}. In high
quality crystals phonons are mostly scattered  by crystalline
defects  and crystal boundaries.
        In the vortex state in  nodal superconductors the quasiparticles  provide another
scattering center. When the thermal phonons  are ballistic, the
phonon scattering due  to the quasiparticles is proportional to
[N(0,$\theta$)]$^2$ at low temperatures \cite{Won_Maki}.
Here $N(0, \theta)$ is the residual density of states in the
presence of a magnetic field and $\theta$ refers to the field
orientation.
Therefore
the $c$ axis phononic thermal conductivity
$\kappa^g$ in Sr$_2$RuO$_4$ is a planar magnetic field is written
\begin{equation}
\frac{\kappa^g}{\kappa_0^g }= [ 1 + \frac{T}{T_0}( 1 + \frac{3\tilde{v}^2
(eH) }{8\pi\Gamma\Delta}\ln(\frac{T}{c\sqrt{eH}} ) ]^{-1}
\end{equation}
for 2D $f$-wave, and
\begin{equation}
\frac{\kappa^g}{\kappa_0^g} = [ 1 + \frac{T}{T_0}( 1 + \frac{\tilde{v}^2
(eH)
}{4\pi\Gamma\Delta}(1-\frac{r}{2}\cos(2\theta))\ln(\frac{T}{c\sqrt{eH}}
) ]^{-1}
\end{equation}
for $d+s$-wave, respectively. Here T$_0$ is a constant of the
dimension of the energy, and $c$ is the phonon velocity. It  is
noteworthy that Eq.(8) does not contain the fourfold term.
\par
When the phononic thermal conductivity dominates as in Sr$_2$RuO$_4$ \cite{Suzuki},
it is more convenient to analyze the thermal resistance
$W=(\kappa^g)^{-1}$
\begin{equation}
\frac{W(H)}{W(0)} - 1 = \frac{T}{T+T_0}
\frac{3\widetilde{v}^2(eH)}{8\pi\Gamma\Delta}\ln(
\frac{T}{c\sqrt{eH}} )
\end{equation}
for 2D $f$-wave. We believe $aH\ln(1/bH)$ dependence is more
consistent with the experiment for $H \parallel [100]$, which indicates the crystal of
Sr$_2$RuO$_4$ is in the clean limit.
This is readily seen from Fig.2.
Also in the clean limit the
specific heat behaves similarly \cite{Suzuki}.
On the other hand for ${\bf H} \parallel [001]$, it appears that the superclean limit appears, ie. linear in $H$ as
shown in Fig. 3.

\begin{figure}
\vspace{-3cm}
     \includegraphics[width=70mm]{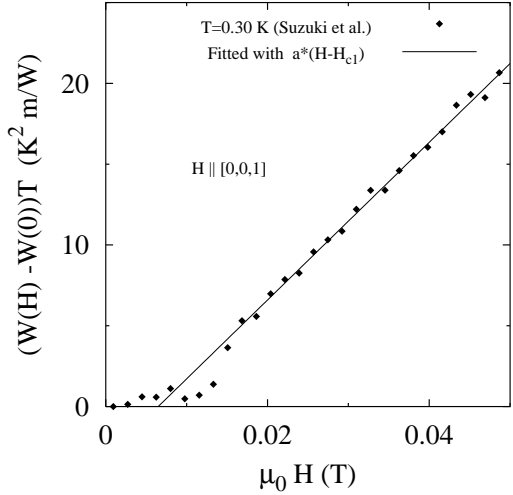}
     \caption{The interplane field dependence of the thermal resistance Ref.\cite{Suzuki} is compared
     with $a*(H-H_{c1})$. }
 \end{figure}

\par
More recently the $b$-axis thermal conductivity of
$\kappa$-(ET)$_2$Cu(NCS)$_2$ in a planar magnetic field is
reported \cite{Izawa}. First of all for $T > 1$ K the thermal
conductivity decreases with increasing $H$ indicating the phonon
dominance. more surprising is an appreciable twofold term($\sim
\cos(2\theta)$) in the thermal conductivity. This suggests
naturally $d+s$-wave model. Indeed we obtain $\Delta({\bf k}) \sim
\cos(2\phi) -0.067$ as already mentioned earlier. As the
temperature is lowered through $T=0.47$ K, there appears a
positive fourfold term. Clearly this is the signature of the
electronic thermal conductivity. Further the magnitude of the
fourfold term is consistent with Eq.(4). Therefore we conclude the
order parameter in $\kappa$-(ET)$_2$Cu(NCS)$_2$ is of $d+s$-wave
with $\Delta({\bf k}) \sim \cos(2\phi) -0.067$. In other words the
nodal lines run  43.08$^\circ$  from the b-axis rather than
45$^\circ$. Since $\kappa$-(ET)$_2$ salts do not have the
tetragonal symmetry, the admixture of $s$-wave component is
allowed. This is somewhat similar  to YBCO, where the tetragonal
symmetry is broken due to the orthorhombic distortion\cite{al-Monod}. This
$d_{x^2-y^2}$-wave is totally unexpected theoretically
\cite{Visentini,Schmalian,Louati,Izawa}.There is a well-known
parallel between high $T_c$ cuprates
and $\kappa$-(ET)$_2$ salts. However, the present result tells
there is a subtle and delicate difference between these two
systems and a more careful study of the pairing interaction is
necessary. This makes the physics in organic superconductors all
the more interesting.

\section{Concluding Remarks}
The gap symmetry is the central issue of new superconductors. We
have shown the angular dependent thermal conductivity in the
vortex state provides a new window to look this question. In this
way we have succeeded in identifying  the gap symmetry of
Sr$_2$RuO$_4$, CeCoIn$_5$ \cite{Izawa_PRL,Petrovic} and
$\kappa$-(ET)$_2$Cu(NCS)$_2$\cite{Izawa}. We expect this method will be
very useful to identify the gap symmetry of $\beta$-(ET)$_2$
salts, $\lambda$-(ET)$_2$ salts and
other organic superconductors and to clarify existing controversy.
Also the success of this method testifies the soundness of both
the BCS theory of nodal superconductors and the Volovik's
semiclassical approach to handle the vortex state in nodal
superconductors.

\end{document}